\documentclass[lettersize,journal]{IEEEtran}
\usepackage{amsmath,amsfonts}
\usepackage{algorithmic}
\usepackage{algorithm}
\usepackage{array}
\usepackage[caption=false,font=normalsize,labelfont=sf,textfont=sf]{subfig}
\usepackage{textcomp}
\usepackage{stfloats}
\usepackage{url}
\usepackage{verbatim}
\usepackage{graphicx}
\usepackage{cite}
\usepackage{tikz}
\usetikzlibrary{arrows.meta,positioning,fit,calc}

\begin{document}
%

\title{Constraining Host-Level Abuse in Self-Hosted Computer-Use Agents via TEE-Backed Isolation}

\author{Di~Lu,~\IEEEmembership{Member,~IEEE,}
Bo Zhang,
Xiyuan Li,
Yongzhi~Liao,
Xuewen~Dong,~\IEEEmembership{Member,~IEEE,}
Yulong~Shen,~\IEEEmembership{Member,~IEEE,}
Zhiquan~Liu, ~\IEEEmembership{Senior Member,~IEEE,}
and~Jianfeng~Ma,~\IEEEmembership{Member,~IEEE}
\thanks{\textbullet\ Di Lu, Bo Zhang, Xiyuan Li, Yongzhi Liao, Xuewen Dong, and Yulong Shen are with the School of Computer Science and Technology, Xidian University, Xi’an, Shaanxi 710071, China, and also with the Shaanxi Key Laboratory of Network and System Security, Xi'an, Shaanxi, 710071, China. 
E-mail: \{dlu, xwdong\}@xidian.edu.cn; z17691248297@gmail.com; lixiyuan9196@outlook.com; liaoyongzhi1010@stu.xidian.edu.cn; ylshen@mail.xidian.edu.cn.\\
\textbullet\ Zhiquan Liu is with the College of Cyber Security, Jinan University, Guangzhou 510632, China. (E-mail: zqliu@vip.qq.com).\\
\textbullet\ Jianfeng Ma is with the School of Cyber Engineering, Shaanxi Key Lab of Network and System Security, Xidian University, Xi’an, Shaanxi, 710071, China. E-mail: jfma@mail.xidian.edu.cn.}%
\thanks{Manuscript received April 19, 2021; revised August 16, 2021.}}

\markboth{IEEE Journal Template,~Vol.~XX, No.~XX, Month~2026}%
{Author \MakeLowercase{\textit{et al.}}: Constraining Host-Level Abuse in Self-Hosted Computer-Use Agents via TEE-Backed Isolation}


\maketitle

\begin{abstract}
Self-hosted computer-use agents (SHCUAs), such as OpenClaw, combine natural-language interaction with direct access to host-side resources, including browsers, files, scripts, system commands, and external communication channels. While useful for automating real tasks, this capability also creates a host-level abuse surface: a legitimately deployed agent may be steered toward unsafe operations through malicious messages, indirect prompt injection, unsafe skills, or tampering along the host-side control path. We argue that such risks cannot be addressed by ad hoc blocking rules alone, because the security criticality of an operation depends jointly on its action type, target object, execution context, and potential effect.

This paper presents an operation-centric model for risk-based confinement of SHCUA operations.
The proposed design keeps ordinary functionality on the constrained REE path, while protecting security-critical classification, authorization, binding, evidence generation, and selected execution-control decisions inside a cloud-native TEE-backed trusted operation plane. We instantiate the architecture on OpenClaw using Intel TDX as the primary trusted backend, with remote terminal-side trusted components verifying TDX-audited commands before constrained local execution. The evaluation shows that the design can block unsafe or policy-disallowed operations before execution, preserve ordinary functionality for allowed workloads, and provide auditable evidence with deployment-dependent overhead.
\end{abstract}


\section{Introduction}
\IEEEPARstart{S}{elf}-hosted computer-use agents (SHCUAs) are emerging as a new class of intelligent systems that combine natural-language interaction with direct access to real host-side tools and resources. Unlike conventional chat assistants that remain confined to text generation, SHCUAs can browse webpages, read and write files, invoke scripts, execute shell commands, call plugins, and communicate through external messaging channels. Their practical value comes from translating high-level user intent into concrete operations over real devices.

OpenClaw is a representative example of this design space. It exposes a combination of properties that makes SHCUAs both practically useful and security-relevant: persistent online availability, multi-channel interaction, extensible tool use, host-local execution, and the ability to orchestrate heterogeneous resources through natural-language control. These properties are not unique to OpenClaw, but OpenClaw provides a concrete and realistic case through which the broader SHCUA security problem can be studied.

The security issue is not simply that SHCUAs possess powerful tools. The deeper problem is that they consolidate planning, reachability, and host action into a single legitimately deployed software entity. Publicly reported cases suggest that OpenClaw has already exposed multiple real-world security problems, including local-instance hijacking via malicious websites (``ClawJacked'')~\cite{oasis2026clawjacked}, one-click remote code execution caused by authentication-token leakage~\cite{github2026openclawrce}, malicious skill poisoning in the skill marketplace~\cite{immersive2026skills}, and malware delivery through fake OpenClaw installers~\cite{malwarebytes2026fakeinstaller}. In the traditional malware model, an attacker typically needs to compromise the host, implant a payload, and maintain persistence before obtaining sustained control. In the SHCUA setting, a substantial part of that capability surface already exists by design. An attacker may therefore attempt to repurpose a legitimate agent through malicious messages, indirect prompt injection, unsafe skills or plugins, browser-delivered content, or tampering with the host-side control path. The concern is not that OpenClaw or similar systems are themselves malware, but that legitimate agent capabilities can be steered toward unsafe host-level effects.

A central challenge is that this abuse surface cannot be described using coarse tool names alone. The same action category may have very different security implications depending on the target object, execution context, and resulting effect. Writing a normal project note and modifying \texttt{/etc/passwd} are both writes, but they differ fundamentally in integrity impact and privilege consequence. Reading a user document and packaging SSH private keys are both reads, but they differ in confidentiality risk and externalization potential. Running a local formatting script and executing a network-fetched shell pipeline are both executions, but they imply different trust assumptions. As a result, SHCUA security cannot be reduced to a flat blacklist of dangerous commands or a binary distinction between benign and sensitive tools.

This observation motivates the first step of our work: an operation-centric security model for SHCUAs. To reason about abuse, policy, audit, and enforcement, the system must represent concrete operation instances in a form that captures the requested action, target object, execution context, and potential effect. Only then can it determine whether an operation is ordinary and task-consistent, whether it should be elevated for stronger handling, or whether it should be denied. Security modeling is therefore the semantic basis on which later authorization, auditing, and verification depend.

Modeling alone is insufficient. Once an operation has been classified, the resulting decision must remain trustworthy during execution. In typical deployments, SHCUAs run inside the host's Rich Execution Environment (REE), together with the operating system, applications, plugins, and browser-delivered content. An attacker who gains influence over this environment may manipulate the control path by altering operation parameters after authorization, suppressing notifications for high-impact actions, or forging audit evidence. In such cases, the system may appear to enforce policy while the actual execution diverges from the approved intent. This reveals a structural limitation of purely REE-based enforcement: the mechanisms intended to constrain abuse may themselves be exposed to manipulation if classification, authorization, operation triggering, and evidence recording all remain in the same environment.

Trusted isolation is therefore needed for the part of the SHCUA control path whose correctness must remain reliable even when the surrounding host environment is not fully trusted. For high-impact operations, security-critical logic such as risk evaluation, authorization, parameter binding, trusted evidence generation, and user-notification triggering should be placed behind a trust boundary stronger than the ordinary REE. The rest of the SHCUA stack, including natural-language interaction, planning, tool selection, and ordinary host-side functionality, can remain on the host side for compatibility and usability.

This separation leads to the main systems tradeoff addressed in this paper. Trusted isolation is valuable but resource-sensitive, and using it as a blanket execution environment for the entire agent would be unnecessarily heavy. We therefore adopt a \emph{risk-driven minimal-confinement} strategy: ordinary low-risk operations remain in the constrained REE path, while only operations and control paths whose security properties cannot be safely preserved in the REE are elevated to a trusted operation plane.

We instantiate this idea in a cloud-native TEE setting, using Intel TDX as the representative backend. The resulting architecture keeps ordinary OpenClaw functionality on the host side, while anchoring security-critical classification, authorization, binding, audit evidence, and selected execution-control decisions in the TEE-backed trusted operation plane. Through this representative case, the paper develops an operation-centric security model and a practical confinement architecture for the broader class of self-hosted computer-use agents.

The main contributions of this paper are as follows:
\begin{itemize}
    \item We identify host-level abuse in SHCUAs as a distinct security problem in which a legitimately deployed agent may be repurposed to perform unsafe operations on real host resources.

    \item We present an operation-centric security model in which the criticality of an SHCUA operation is determined jointly by action type, target object, execution context, and potential effect, thereby providing a principled basis for audit, verification, and enforcement.

    \item We establish the role of trusted isolation for security-critical SHCUA operations, showing that classification, authorization, binding, evidence generation, and selected execution-control decisions cannot be safely entrusted to the REE alone.

    \item We derive a risk-driven minimal-confinement framework and a cloud-native TEE-backed trusted operation architecture that constrains security-critical host operations while preserving ordinary SHCUA usability.

    \item We implement the proposed design using OpenClaw as a representative SHCUA and evaluate it in a TDX-centered deployment with heterogeneous remote terminal verification.
\end{itemize}

The remainder of this paper is organized as follows. Section~II introduces the background, threat model, and scope assumptions for SHCUAs. Section~III presents the security modeling of SHCUA operations, including operation instances, risk factors, risk assessment, and enforcement decisions. Section~IV develops the system design, covering trusted operation requests, isolation-based enforcement, audit and verifiable evidence, and user notification and confirmation. Section~V describes the adaptation of the proposed design to cloud-native TEEs, using Intel TDX as the representative backend. Section~VI presents the evaluation, including the experimental setup, performance evaluation, and security analysis. Section~VII reviews related work, and Section~VIII concludes the paper.

\section{Background, Threat Model and Assumptions}

\subsection{Self-Hosted Computer-Use Agents}
Self-Hosted Computer-Use Agents (SHCUAs) represent an emerging class of AI agents that are deployed and operated by users on their own endpoints or infrastructure. Unlike conventional assistants that remain primarily confined to text generation, SHCUAs are designed to interact with real host-side resources through browser automation, file access, script execution, command invocation, plugin use, and external communication channels. Their defining characteristic is that they translate natural-language inputs into concrete operations over real system resources. This capability makes them practically useful in everyday workflows, but it also changes their security significance. Once an agent can not only interpret user intent but also act on the host environment, it effectively acquires operational influence over local files, applications, services, and communication paths. As a result, the security problem is no longer limited to incorrect responses or unsafe suggestions, but extends to whether a legitimately deployed agent can be induced to perform unsafe host-level operations. This shift from conversational risk to operational risk is what makes SHCUAs fundamentally different from earlier assistant models.

\subsection{Isolated Execution Environments}
A variety of isolated execution environments are available today for constraining software behavior and protecting security-critical operations. At one end, application sandboxes and process-level confinement mechanisms can restrict filesystem access, system calls, and network communication with relatively low deployment cost, but they typically rely on the correctness and trustworthiness of the host operating system. Containers provide stronger packaging and namespace-based isolation, and are widely deployable across PCs, servers, and Linux-capable embedded devices, yet they still fundamentally execute within the host kernel and therefore inherit its security assumptions. Virtual machines provide stronger isolation by separating guest execution from the host via a hypervisor, at the cost of greater resource overhead. Trusted Execution Environments (TEEs) provide hardware-backed protection for selected code and data, enabling stronger confidentiality and integrity guarantees for security-critical logic, though usually under tighter interface and resource constraints. In practice, these mechanisms differ in isolation strength, deployment cost, performance overhead, and trust assumptions. For SHCUAs, isolated execution environments thus form an important technical basis for later mechanisms that constrain agent behavior and protect the trustworthiness of auditing and verification.

\subsection{Threat Model}
We consider an adversary whose goal is to induce an SHCUA to perform unsafe host-level operations over real system resources. Unlike traditional malware-oriented threat models, the adversary does not necessarily need to compromise the host first. Instead, the adversary may influence the agent through malicious messages, browser-delivered content, prompt injection, unsafe plugins, tampering with the host-side control path, or interference with the model dependency chain, such as poisoned model updates, manipulated inference services, or untrusted model-side inputs that alter the agent’s decision process. Under this model, the SHCUA is a legitimately deployed software entity that already possesses substantial operational reach over local files, applications, services, and communication channels. The primary security risk is therefore not merely unauthorized access to the agent itself, but the possibility that its existing capabilities are repurposed toward unsafe effects by manipulating the decision logic and control flow on which those capabilities depend. In particular, an attacker may attempt to interfere with host-side security-relevant handling so that a sensitive request is treated as if it were an ordinary one, or so that authorization, notification, auditing, or evidence recording no longer faithfully reflects the actual operation being performed.

\subsection{Assumptions and Scope}

To make the problem well-defined, we adopt the following assumptions about deployment and attack scope. First, we assume that the SHCUA is deployed within a constrained runtime domain rather than as an unrestricted host process, so that it does not directly hold all privileges required for security-critical host interactions. Second, we assume that the REE cannot be treated as fully trustworthy for security-critical decision-making, authorization, audit, or evidence generation. Purely REE-side protection is therefore insufficient for operations whose safety depends on trustworthy classification, scope binding, and evidence semantics.

This paper focuses on cloud-native SHCUA deployment, where a server-class platform hosts the SHCUA runtime and a TEE-backed trusted operation plane protects the security-critical control path. We use Intel TDX as the representative backend in our prototype. Remote terminal devices may appear as controlled endpoints, but they are not considered as direct hosting platforms for the full SHCUA runtime or for the complete trusted operation plane. In our evaluation, their local trusted components are used only to verify TDX-audited commands before constrained local execution.

We do not consider physical attacks, direct compromise of the underlying TEE substrate, or harmful operations that are explicitly requested by an authorized user and permitted by policy. Our focus is on modeling, constraining, auditing, and verifying security-relevant SHCUA operations under a cloud-native trusted-control architecture, rather than on securing every layer of the surrounding software, hardware, or remote-terminal stack.

\section{Security Modeling of SHCUA Operations}

\subsection{Modeling Objective}
The security problem of SHCUAs cannot be adequately assessed based on coarse tool names or interface categories alone. The same nominal action may have radically different security implications depending on the object it targets, the context in which it is requested, and the effect it may produce. For example, a file-write action may correspond to ordinary workspace editing in one case and to privileged configuration modification in another; a read request may refer to an innocuous project document or to highly sensitive credentials. As a result, SHCUA security requires a principled model of concrete operational instances rather than ad hoc reasoning based on commands, tools, or prompts alone.

The purpose of this model is not merely descriptive. It must provide a unified basis for deciding whether an operation is ordinary or security-critical, whether it may remain in the REE or should be elevated for stronger control, what should be audited and verified, and what evidence should later be produced for accountability. In this sense, the security model serves as the semantic foundation for later confinement, auditing, and verification.

Formally, our objective is to define: 1) a structured representation of SHCUA operation instances; 2) a set of risk projections that map actions, objects, contexts, and effects into comparable security levels; 3) a risk aggregation function that assigns each operation instance a security-criticality level; and 4) a decision function that maps the resulting level to a concrete enforcement decision.

\subsection{Operation Instance}
Let $\mathcal{S}$, $\mathcal{A}$, $\mathcal{R}$, $\mathcal{C}$, and $\mathcal{E}$ denote the domains of subjects, actions, target objects, execution contexts, and potential effects, respectively. A concrete SHCUA operation instance is modeled as
\begin{equation}
\mathcal{O} = \langle s, a, r, c, e \rangle \in \mathbb{O},
\end{equation}
where
\begin{equation}
\mathbb{O} = \mathcal{S} \times \mathcal{A} \times \mathcal{R} \times \mathcal{C} \times \mathcal{E}.
\end{equation}

Here, $s \in \mathcal{S}$ denotes the effective subject that initiates the operation, $a \in \mathcal{A}$ the requested action, $r \in \mathcal{R}$ the target object, $c \in \mathcal{C}$ the execution context, and $e \in \mathcal{E}$ the potential effect. This representation makes explicit that the security meaning of an SHCUA operation is determined jointly by who requests it, what is to be done, to what target, under which circumstances, and with what possible outcome.

The subject $s$ should be understood as the effective initiator as seen by the system, rather than only a human user in the narrow sense. It may correspond to an SHCUA session, a tool-invocation chain, or a request whose behavior is influenced by external content. The action component $a$ abstracts the intended operation category, such as read, write, execute, send, invoke, modify, or configure. The object component $r$ refers to the resource on which the action operates, such as a file, directory, credential, configuration state, service endpoint, command target, or external destination. The context component $c$ captures the situational information needed for security interpretation, while $e$ represents the potential security-relevant effect if the operation is carried out.

\subsection{Operation Risk Factors}
The security criticality of an SHCUA operation depends on several risk factors. We model these factors through four ordered projections over the operation instance.

Let
\begin{equation}
L_A,\; L_R,\; L_C,\; L_E
\end{equation}
denote ordered level sets for action sensitivity, object criticality, contextual risk, and effect severity, respectively. In the simplest case, these sets may be instantiated as ordinal levels such as
\begin{equation}
L_A = L_R = L_C = L_E = \{0,1,2,3\},
\end{equation}
where larger values indicate stronger security significance.

We then define the following projection functions:
\begin{align}
\alpha &: \mathcal{A} \rightarrow L_A, \\
\beta  &: \mathcal{R} \rightarrow L_R, \\
\gamma &: \mathcal{C} \rightarrow L_C, \\
\delta &: \mathcal{E} \rightarrow L_E.
\end{align}

The function $\alpha(a)$ captures the baseline sensitivity implied by the action type. For example, executing code, modifying persistent configuration, or sending data across a trust boundary generally carries stronger security implications than reading an ordinary workspace file.

The function $\beta(r)$ captures the sensitivity and criticality of the target object. Objects differ substantially in their security significance. Public or ordinary user files differ from sensitive credentials, privileged configuration files, persistent control-plane state, and other resources whose modification or disclosure may have system-level consequences. For practical reasoning, target objects may be classified into categories such as public, ordinary, sensitive, and critical.

The function $\gamma(c)$ captures the security relevance of the execution context. Relevant contextual factors include request origin, whether the request is browser-derived or plugin-derived, whether the requested action is consistent with the current task, whether the user is present for oversight, whether the operation crosses a trust boundary, and whether it participates in a multi-step chain whose combined effect is more dangerous than any single step in isolation.

The function $\delta(e)$ captures the severity of the potential effect if the operation is executed. In this paper, we focus on effects such as confidentiality loss, integrity violation, availability impact, privilege or control amplification, and the externalization of protected resources.

For any operation instance $\mathcal{O} = \langle s,a,r,c,e\rangle$, these projections induce the risk feature vector
\begin{equation}
\mathbf{v}(\mathcal{O}) = \langle \alpha(a), \beta(r), \gamma(c), \delta(e) \rangle.
\end{equation}

Taken together, these risk factors prevent the model from reducing security reasoning to crude labels such as ``read,'' ``write,'' or ``execute.'' Instead, they enable distinguishing between operation instances that are superficially similar but materially different in security consequences.

\subsection{Risk Assessment and Security Levels}
Based on the factors above, we define the security criticality of an operation instance through a risk aggregation function
\begin{equation}
\rho : \mathbb{O} \rightarrow L,
\end{equation}
where $L$ is an ordered set of security levels. Concretely,
\begin{equation}
\rho(\mathcal{O}) = \Phi\big(\alpha(a), \beta(r), \gamma(c), \delta(e)\big),
\end{equation}
for $\mathcal{O} = \langle s,a,r,c,e\rangle$.

Here, $\Phi$ is a deployment-dependent aggregation function that combines baseline action sensitivity, target-object criticality, contextual risk, and effect severity into a single security-criticality level. The model is not intended to define a universal risk calculus for all SHCUA deployments. Instead, it provides a policy-computable abstraction: concrete systems may instantiate $\Phi$ using policy tables, rule sets, or deployment-specific risk classifiers, as long as the resulting level remains bound to the concrete operation instance and can be consumed by the later enforcement path.

For practical enforcement, the codomain $L$ may be instantiated as an ordered level set such as
\begin{equation}
L = \{L_0, L_1, L_2, L_3\},
\end{equation}
where:
\begin{itemize}
    \item $L_0$ denotes routine operations;
    \item $L_1$ denotes low-to-moderate security relevance;
    \item $L_2$ denotes high security relevance; and
    \item $L_3$ denotes security-critical operations.
\end{itemize}

These levels are not used merely to rank operations qualitatively. They determine what degree of control, auditing, trusted handling, and verification each operation requires, and they provide the input to the enforcement decision function defined in the next subsection.

\subsection{Enforcement Decisions}
The assessed risk level of an operation instance determines the corresponding enforcement decision. Let
\begin{equation}
\mathcal{D} = \{d_{\mathrm{ree}}, d_{\mathrm{ia}}, d_{\mathrm{ie}}, d_{\mathrm{uc}}, d_{\mathrm{deny}}\}
\end{equation}
denote the decision set, where:
\begin{itemize}
    \item $d_{\mathrm{ree}}$ denotes direct execution in the REE;
    \item $d_{\mathrm{ia}}$ denotes isolated authorization with constrained execution;
    \item $d_{\mathrm{ie}}$ denotes isolated execution;
    \item $d_{\mathrm{uc}}$ denotes isolated execution with user notification or confirmation; and
    \item $d_{\mathrm{deny}}$ denotes denial.
\end{itemize}

We define an enforcement decision function
\begin{equation}
\eta : \mathbb{O} \rightarrow \mathcal{D},
\end{equation}
which maps each operation instance to a concrete enforcement outcome. In the simplest case, $\eta$ can be defined over the assessed risk level. Recall from the previous subsection that $\rho(\mathcal{O}) \in L$ denotes the security-criticality level assigned to operation instance $\mathcal{O}$ by the risk aggregation function. We then define
\begin{equation}
\eta(\mathcal{O}) = \Psi(\rho(\mathcal{O}), \chi(\mathcal{O})),
\end{equation}
where $\chi(\mathcal{O})$ captures auxiliary policy conditions, such as whether user confirmation is possible, whether a trusted-isolation backend is available, or whether the target object belongs to a mandatory protected class.

A typical policy may be expressed as
\begin{equation}
\eta(\mathcal{O}) =
\begin{cases}
d_{\mathrm{ree}}, & \rho(\mathcal{O}) = L_0,\\
d_{\mathrm{ia}},  & \rho(\mathcal{O}) = L_1,\\
d_{\mathrm{ie}},  & \rho(\mathcal{O}) = L_2,\\
d_{\mathrm{uc}},  & \rho(\mathcal{O}) = L_3 \land \chi(\mathcal{O}) = 1,\\
d_{\mathrm{deny}},& \rho(\mathcal{O}) = L_3 \land \chi(\mathcal{O}) = 0.
\end{cases}
\end{equation}

To explicitly characterize when trusted isolation becomes necessary, we further define a predicate
\begin{equation}
\mathsf{NeedIso} : \mathbb{O} \rightarrow \{0,1\},
\end{equation}
such that
\begin{equation}
\mathsf{NeedIso}(\mathcal{O}) =
\begin{cases}
1, & \eta(\mathcal{O}) \in \{d_{\mathrm{ia}}, d_{\mathrm{ie}}, d_{\mathrm{uc}}, d_{\mathrm{deny}}\},\\
0, & \eta(\mathcal{O}) = d_{\mathrm{ree}}.
\end{cases}
\end{equation}

Thus, the operation model does not terminate at classification; it directly drives how the system constrains SHCUA behavior and what trust guarantees are required of the execution path. More importantly, it provides a unified basis for auditing and verification. Because operations are evaluated at the level of concrete instances, the system can later verify not only whether an operation occurred, but also whether it was appropriately classified, whether the corresponding enforcement decision was justified, and whether the produced evidence matches the claimed authorization and execution path. This is precisely the bridge from security modeling to the later trusted-isolation design: only with an operation-centric view can the system decide which SHCUA behaviors remain ordinary software activity and which become security-critical actions requiring stronger trust guarantees.

\subsection{Running Example}
To make the model concrete, we consider a simple SHCUA-assisted document-management task. A user asks the SHCUA to organize project materials in a workspace, summarize selected notes, and prepare a result file for later sharing. Under this task, the SHCUA may perform ordinary operations such as reading project notes from \texttt{/workspace/notes.md} and writing a summary to \texttt{/workspace/summary.txt}. These actions are task-consistent, target ordinary workspace objects, and usually imply limited security impact.

For example, consider the write operation
\begin{equation}
\mathcal{O}_1 = \langle s_1,\; \texttt{write},\; \texttt{/workspace/summary.txt},\; c_1,\; e_1 \rangle.
\end{equation}
If the context $c_1$ indicates that the request is consistent with the current document-management task, and the effect $e_1$ is limited to ordinary workspace modification, then the corresponding risk feature vector may be written as
\begin{equation}
\mathbf{v}(\mathcal{O}_1)
=
\langle \alpha(\texttt{write}),\beta(\texttt{/workspace/summary.txt}),\gamma(c_1),\delta(e_1)\rangle
=
\langle 1,1,0,1\rangle.
\end{equation}
Accordingly, the risk level may be assessed as
\begin{equation}
\rho(\mathcal{O}_1)=L_0,
\end{equation}
which leads to
\begin{equation}
\eta(\mathcal{O}_1)=d_{\mathrm{ree}}.
\end{equation}

Now consider a superficially similar operation in which the SHCUA attempts to modify a privileged system file:
\begin{equation}
\mathcal{O}_2 = \langle s_2,\; \texttt{write},\; \texttt{/etc/passwd},\; c_2,\; e_2 \rangle.
\end{equation}
Although $\mathcal{O}_2$ has the same nominal action type as $\mathcal{O}_1$, it targets a critical object and may cause integrity compromise or privilege escalation. Its corresponding risk feature vector may therefore be written as
\begin{equation}
\mathbf{v}(\mathcal{O}_2)
=
\langle \alpha(\texttt{write}),\beta(\texttt{/etc/passwd}),\gamma(c_2),\delta(e_2)\rangle
=
\langle 1,3,2,3\rangle.
\end{equation}
The aggregated risk level may then be assessed as
\begin{equation}
\rho(\mathcal{O}_2)=L_3,
\end{equation}
which in turn leads to a strong enforcement decision such as
\begin{equation}
\eta(\mathcal{O}_2)\in \{d_{\mathrm{uc}}, d_{\mathrm{deny}}\}.
\end{equation}

A similar contrast applies to read operations. Reading an ordinary project note and reading \texttt{\~{}/.ssh/id\_rsa} are both nominally \texttt{read} actions, yet they differ fundamentally in target-object criticality and potential effect. This running example illustrates how the proposed model evaluates concrete SHCUA behaviors through the full chain from operation instance to risk factors, from risk level to enforcement decision. It also provides a basis for the evaluation in later sections, where the consistency, effectiveness, and overhead of model-driven enforcement can be measured on representative SHCUA tasks.

\section{System Design}

\subsection{Design Overview}
The proposed design is based on a simple observation: constraining SHCUA abuse requires more than attaching a security monitor to an otherwise unrestricted agent process. Instead, the SHCUA must first be hosted inside a constrained REE-side runtime domain so that it does not directly possess all privileges required for security-critical host interactions. This constrained deployment establishes the basis for non-bypassable control, while trusted isolation protects the small subset of control and evidence paths that cannot be safely entrusted to the REE alone.
Constrained hosting therefore addresses where the SHCUA may execute and what direct authority it is denied, whereas trusted isolation addresses who can be trusted to classify, authorize, audit, and later verify security-critical operations once they are requested.
In this sense, constrained deployment provides controllability, whereas trusted isolation provides trustworthiness for security-critical classification, authorization, parameter binding, auditing, and verification.

Guided by this distinction, our design follows a risk-driven minimal-confinement philosophy. The SHCUA itself remains primarily on the REE side in order to preserve compatibility and efficiency for ordinary functionality, while only the minimum security-critical control logic is elevated into trusted isolation. Concretely, the SHCUA runtime produces concrete operation requests during task execution; these are first extracted on the REE side by the operation-extraction layer and then normalized by the dispatcher and request builder into trusted operation requests. The trusted operation plane interprets each such request as a modeled operation $\mathcal{O}$ and carries out the mapping $\mathcal{O}\mapsto \rho(\mathcal{O}) \mapsto \eta(\mathcal{O})$ under trusted isolation. Low-risk operations may then continue through the ordinary REE execution path under the resulting trusted decision, whereas security-critical operations remain on the trusted path and are handled through trusted authorization, binding, audit, and notification control.

\begin{figure}[!t]
\centering
\includegraphics[width=3.5in]{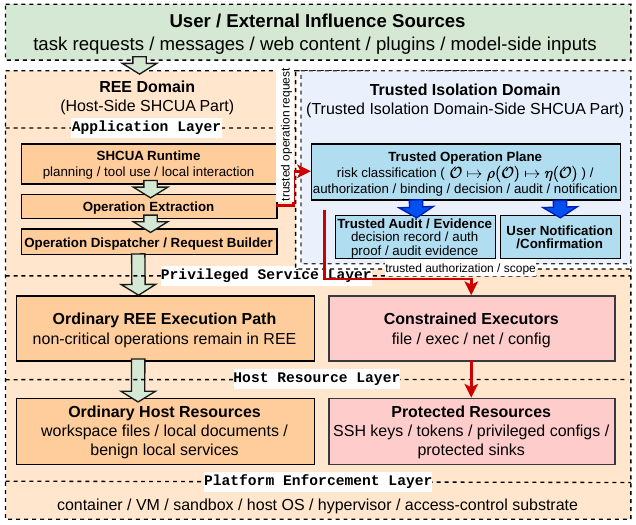}
\caption{Overview of the proposed SHCUA protection architecture. Low-risk operations remain on the constrained REE path, while security-critical operations are forwarded as trusted operation requests to a minimal trusted operation plane for risk classification, authorization, auditing, and notification control before protected resources are accessed through constrained executors.}
\label{fig:system_overview}
\end{figure}

Fig.~\ref{fig:system_overview} shows the resulting architecture. On the REE side, SHCUA-related components are organized into application, privileged-service, host-resource, and platform-enforcement layers. On the trusted side, the trusted isolation domain contains only a minimal trusted operation plane responsible for risk classification, authorization, binding, decision, trusted audit and evidence generation, and user-notification or confirmation triggering. This architecture deliberately avoids migrating the full SHCUA stack into trusted isolation. Instead, it isolates only the minimal control-plane logic whose correctness, auditability, and verifiability cannot be guaranteed by the REE alone.
Constrained deployment limits the SHCUA's direct authority over host resources, while trusted isolation ensures that the control decisions governing security-critical operations remain trustworthy even when the surrounding REE does not.
The following subsections detail the trusted operation request format, the isolation-based enforcement flow, the trusted auditing and evidence mechanism, and the handling of user notification and confirmation.

\subsection{Trusted Operation Requests}
In the architecture of Fig.~\ref{fig:system_overview}, the trusted operation request is the interface object that connects the REE-side analysis path to the trusted operation plane. Once the SHCUA runtime emits a concrete operation, the operation-extraction layer captures the relevant action, object, and contextual information, and the operation dispatcher and request builder construct a normalized request for trusted handling. The modeled representation $\mathcal{O}$, together with the corresponding risk level $\rho(\mathcal{O})$ and enforcement decision $\eta(\mathcal{O})$, is then derived inside the trusted operation plane. In this way, the decision as to whether the operation may remain on the ordinary REE path or must continue on the trusted path is itself made under trusted isolation. For the latter case, the dispatcher submits the trusted operation request as the canonical input for subsequent trusted handling.

Formally, the trusted operation request is represented as
\begin{equation}
A = \langle sid, act, obj, scope, ctx, level, seq, ttl \rangle,
\end{equation}
where $sid$ denotes the session identifier, $act$ the requested action, $obj$ the target object, $scope$ the authorized operational boundary, $ctx$ the bound execution context, $level$ the assessed security level, $seq$ the request sequence number, and $ttl$ the request lifetime. In system terms, $A$ is not another abstract security model, but the normalized runtime control object that leaves the REE-side request-building path and enters the trusted operation plane.

The role of $A$ is to carry exactly the information the trusted operation plane needs to make a security-critical decision without depending on tool-specific REE semantics. The pair $(act,obj)$ identifies the requested operation and its target. The field $scope$ encodes the execution boundary that later constrains the behavior of REE-side privileged services and constrained executors. The field $ctx$ binds the request to the relevant execution state, while $level$ conveys the result of the trusted risk assessment. Finally, $(seq,ttl)$ support request freshness, replay resistance, and request--response consistency. Together, these fields provide the minimum control information required for trusted authorization, parameter binding, audit-evidence generation, and later verification.

From the architectural perspective, the trusted operation request serves as the key normalization boundary in the system. On the left side of Fig.~\ref{fig:system_overview}, SHCUA behaviors may originate from heterogeneous tools, plugins, browser actions, or other host-side interactions. On the right side, however, the trusted operation plane should not be required to understand every tool-specific interface separately. By converting security-relevant operations into the uniform request form $A$, the request builder decouples REE-side behavioral diversity from trusted-side security control. 
This also makes the trusted control path less dependent on tool-specific host-side interfaces: the TEE-backed trusted operation plane consumes a uniform request object, while REE-side components remain responsible for adapting concrete tools, commands, and resource accesses to that interface.

Operationally, a trusted operation request is generated whenever the operation must be submitted to the trusted operation plane for the trusted mapping from $\mathcal{O}$ to $\rho(\mathcal{O})$ and then to $\eta(\mathcal{O})$. In this sense, $A$ is the concrete runtime object that carries the operation-centric security model into the later system pipeline. The next subsections build directly on this interface: isolation-based enforcement uses $A$ to perform trusted authorization and scope-constrained execution, trusted auditing binds evidence to $A$ and its decision outcome, and user-notification logic uses $A$ to determine when high-impact operations must be surfaced to the user.

\subsection{Isolation-Based Enforcement}
Once a trusted operation request $A$ has been constructed, the system enforces it through the trusted path. The trusted operation plane first validates the request and its binding to the current session and context, then performs the trusted mapping $\mathcal{O}\mapsto \rho(\mathcal{O}) \mapsto \eta(\mathcal{O})$ under the applicable policy constraints, and finally determines the corresponding enforcement state. Depending on the result, the operation may be allowed under isolated authorization, elevated to isolated execution, subjected to user notification or confirmation, or denied.
This step remains necessary even when the SHCUA is already deployed inside a constrained container or virtual machine. Such constrained hosting can prevent the SHCUA from directly exercising unrestricted host authority, but it does not by itself make the subsequent classification, authorization, scope binding, or execution approval trustworthy. If these security-critical decisions were still left entirely to the REE, a malicious or compromised host runtime could tamper with the decision process or misrepresent its outcome. The trusted operation plane is therefore introduced not to repeat the function of deployment-time confinement, but to provide a trustworthy control point for security-critical classification and enforcement.

Actual interaction with protected resources is not performed by the trusted operation plane itself, but by constrained executors on the REE side. These executors are privileged service components that mediate access to security-relevant files, configurations, commands, and protected communication endpoints. Their role is to ensure that security-critical host interactions cannot be performed directly by the SHCUA runtime, but only through a constrained host-side path governed by the authorization and scope determined by the trusted operation plane. In this way, trusted isolation protects the security-critical classification, decision, and binding logic, while constrained executors serve as bounded host-side execution agents for operations whose target resources remain outside the trusted isolation domain.

To support this bounded execution model, the trusted operation plane returns an authorization result bound to the request content, the approved scope, and the relevant execution context. The constrained executor verifies this result before accessing the corresponding protected resource. As a consequence, a malicious or compromised REE cannot broaden an approved operation into a different one through the normal constrained-execution path without violating the trusted authorization, approved scope, or the consistency of the resulting evidence chain. At the same time, the design preserves risk-driven minimal confinement: ordinary low-risk operations remain on the REE path, while only security-critical operations are redirected to the trusted path. This does not mean that REE-side executors are themselves fully trusted. Rather, the design recognizes that protected host resources remain outside the trusted isolation domain and therefore must still be accessed through host-side mechanisms. The role of the trusted operation plane is to ensure that security-critical classification, authorization, binding, and evidence generation remain trustworthy, while any deviation by a compromised executor is forced to manifest as an authorization or evidence inconsistency, or as a denial-of-service-style failure, rather than as an undetectable corruption of the trusted control logic itself.

\subsection{Audit and Verifiable Evidence}
Auditability is a first-class goal of the proposed design rather than a side effect of execution logging. In the SHCUA setting, security-relevant operations may traverse multiple REE-side components before reaching protected resources. If audit records were generated only by ordinary REE software, a malicious or compromised REE could suppress, alter, reorder, or forge them. For this reason, the architecture places the generation of security-critical audit evidence inside the trusted operation plane rather than relying solely on host-side logs.
In other words, deployment-time confinement can limit what the SHCUA may directly reach, but it cannot by itself guarantee that later classification outcomes, authorization results, audit records, and evidence semantics are generated by a trustworthy party.

For each trusted operation request $A$, the trusted operation plane produces a corresponding evidence record
\begin{equation}
E = \langle sid, act, obj, scope, level, seq, dec, ts, res \rangle,
\end{equation}
where $sid$ denotes the session identifier, $act$ the requested action, $obj$ the target object, $scope$ the approved operational boundary, $level$ the assessed security level, $seq$ the request sequence number, $dec$ the enforcement decision, $ts$ the trusted timestamp or trusted event marker, and $res$ the execution result or status. Intuitively, $E$ binds together what was requested, how it was classified, what decision was made, under what scope it was authorized, and how execution was eventually completed.

The evidence record supports audit reconstruction and consistency checking across the request, decision, and result. Later verification can examine whether the executed operation matches the original trusted request, whether the decision recorded in $E$ is consistent with the risk level carried by $A$, and whether the reported result remains within the approved scope. Thus, the evidence record is not merely a log entry; it is a compact statement of authorization and execution semantics produced by the trusted path.

Operationally, evidence generation is triggered whenever an operation enters the trusted path. After receiving a request $A$, the trusted operation plane records the classification outcome, enforcement decision, and approved scope before execution proceeds. Once the corresponding constrained executor completes or rejects the requested operation, the outcome is returned and incorporated into the final evidence record. This creates a verifiable linkage between request, decision, and result, rather than leaving auditability to disconnected host-side traces.

This design does not require that all data storage itself reside inside trusted isolation. Instead, what must remain trustworthy is the generation and binding of the evidence semantics. The evidence record may later be exported, sealed, or stored by the surrounding system, but its security value derives from the fact that the critical fields are produced by the trusted operation plane and tied to the trusted request and decision path. In this way, the architecture supports later accountability, inconsistency detection, and policy-level verification even when the REE cannot be treated as fully trustworthy.

\subsection{User Notification and Confirmation}
Not all security-critical operations should be executed automatically, even after they have been classified and routed through the trusted path. For selected high-impact operations, the system must additionally surface the intended action to the user before execution. Typical cases include operations that access highly sensitive resources, cross trust boundaries with protected data, or may cause irreversible changes to the host state.

In the proposed architecture, user notifications and confirmations are triggered by the trusted operation plane rather than by REE logic alone. Once a trusted operation request $A$ is assessed as requiring user involvement, the trusted plane generates a notification decision bound to the request identity, the approved scope, and the current security level. This ensures that the user is notified about the specific security-relevant operation under consideration, rather than about an unbound or potentially mutable REE-side interpretation of it.

If confirmation is required, execution does not proceed until the corresponding confirmation result is returned and validated against the original request context. If the user approves, the operation may continue under the previously determined authorization and scope constraints; otherwise, it is denied. In this way, user confirmation is treated not as an optional interface prompt, but as a trusted control step in the enforcement chain.

This mechanism preserves user visibility for high-impact operations and binds user involvement to the same trusted request and decision path used for classification, authorization, auditing, and execution. As a result, notification and confirmation become part of the verifiable security chain rather than an optional usability feature attached outside it.

\section{Adaptation to Cloud-Native TEEs}

This section describes how the proposed protection architecture is adapted to a cloud-native TEE setting, using Intel TDX as the representative backend. The same logic also applies to similar VM-based confidential-computing platforms such as AMD SEV. We focus on this setting because practical SHCUA systems such as OpenClaw depend on server-class resources for planning, tool orchestration, policy handling, and audit generation; the cloud-native TEE therefore hosts the trusted operation plane, while ordinary SHCUA functionality remains on the host side.

\begin{figure}[!t]
\centering
\includegraphics[width=3.5in]{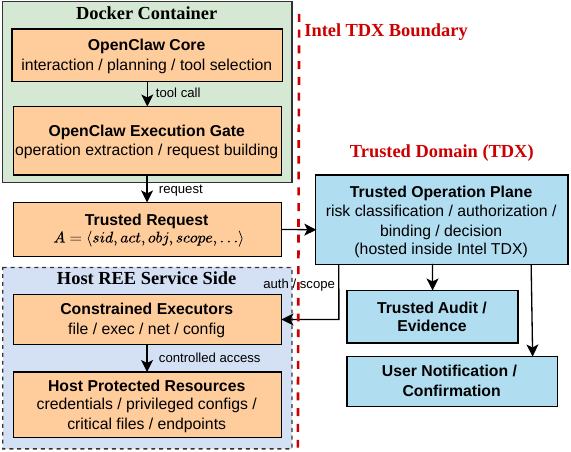}
\caption{Deployment-oriented adaptation in a cloud-native TEE setting. OpenClaw runs inside a Docker container on the REE side, while the trusted operation plane is hosted inside Intel TDX. REE-side constrained executors access protected host resources only under scoped trusted authorization.}
\label{fig:openclaw_tee_adaptation}
\end{figure}

\subsection{Cloud-Native Deployment Model}

In the cloud-native deployment shown in Fig.~\ref{fig:openclaw_tee_adaptation}, OpenClaw runs inside a Docker container on the REE side. This container preserves ordinary OpenClaw functionality, including user interaction, task planning, tool selection, file access, command invocation, and communication with host-side services. It also serves as the constrained runtime boundary that prevents OpenClaw from directly exercising unrestricted authority over protected host resources.

Security-relevant tool calls are intercepted by the OpenClaw execution gate before they reach privileged host-side services. The execution gate extracts the requested operation and normalizes it into a trusted operation request, which is then forwarded to the TDX-backed trusted operation plane. The trusted plane performs security-critical classification, authorization, binding, audit, and notification-related decisions inside the cloud-native TEE domain.

This deployment follows the risk-driven minimal-confinement principle. Ordinary low-risk operations remain on the constrained REE path, while operations requiring trusted classification, authorization, scope binding, audit evidence, or user notification are elevated to the trusted operation plane. In this way, the architecture keeps the OpenClaw runtime usable and compatible with host-side tools, while placing the security-critical control path inside Intel TDX.

\subsection{TEE-Backed Trusted Operation Plane}

The trusted operation plane inside the cloud-native TEE is responsible for executing the security-critical part of the operation model defined in Section~III. After receiving a trusted operation request, it reconstructs the corresponding operation instance $\mathcal{O}=\langle s,a,r,c,e\rangle$ from the request fields, including the effective subject, requested action, target object, bound execution context, and potential effect. It then derives the risk feature vector $v(\mathcal{O})$, computes the security level $\rho(\mathcal{O})$, and obtains the enforcement decision $\eta(\mathcal{O})$ under the applicable policy.

Concretely, the trusted operation plane performs four main functions. First, it evaluates the operation-level risk factors $\alpha(a)$, $\beta(r)$, $\gamma(c)$, and $\delta(e)$, so that the requested action is interpreted together with its target object, execution context, and possible effect rather than by tool name alone. Second, it applies the risk aggregation function $\Phi$ to obtain $\rho(\mathcal{O})$ and derives the corresponding enforcement decision through $\eta(\mathcal{O})=\Psi(\rho(\mathcal{O}),\chi(\mathcal{O}))$. This prevents a compromised or manipulated REE-side component from unilaterally treating a sensitive operation as an ordinary one. Third, it binds the resulting authorization to the concrete action, object, scope, context, sequence number, and freshness information carried by the trusted request. Fourth, it generates trusted audit evidence and triggers user notification or confirmation when required by the resulting decision.

The output of the trusted operation plane is therefore not a generic permission to use a tool. It is a scoped authorization result bound to a specific operation instance and to the model-derived decision path from $\mathcal{O}$ to $\rho(\mathcal{O})$ and then to $\eta(\mathcal{O})$. This binding is essential because the actual protected resources still reside outside the trusted domain and are accessed through host-side constrained executors. Without such binding, a compromised host-side component could attempt to broaden an approved operation, redirect it to a different target, or reuse an authorization result in a different context.

\subsection{REE-Side Constrained Execution}

After the trusted operation plane returns a scoped authorization result, the corresponding host-side action is carried out by an REE-side constrained executor. The executor does not make the security decision itself; it only checks whether the pending action matches the trusted authorization in terms of action type, target object, approved scope, context, sequence number, and freshness.

If the check succeeds, the executor dispatches the operation to the protected host resource, such as a file, configuration object, command interface, or communication sink. If the check fails, the operation is rejected or reported as an inconsistency. This keeps host-resource access compatible with ordinary system interfaces while preventing the OpenClaw runtime from directly broadening, redirecting, or reusing a trusted authorization.

\subsection{Security Implications and Limitations}

This adaptation anchors security-critical classification, authorization, binding, audit evidence, and selected user-visible control decisions inside the cloud-native TEE. As a result, the REE-side OpenClaw runtime is no longer the sole authority over sensitive host operations, and constrained executors can rely on scoped trusted authorization rather than mutable host-side decisions or logs.

The design does not make the entire OpenClaw stack or host execution path trusted. The REE may still deny service, drop requests, interfere with low-risk operations, or refuse to execute an authorized action. The design also assumes that the cloud-native TEE backend and the trusted operation plane are correctly implemented and not directly compromised. Since protected resources remain outside the trusted domain, deviations by compromised host-side components are handled through denial, inconsistency detection, or audit-evidence mismatch, rather than by eliminating all REE-side risk.

\section{Evaluation}

\subsection{Experimental Setup}

\subsubsection{Platform Configuration}

Our evaluation follows a typical server-plus-remote-terminal IoT deployment scenario. A more capable Intel TDX server hosts the SHCUA runtime, performs the primary trusted handling of security-relevant operations, and produces audited control commands. Heterogeneous terminal devices, including an ARM TrustZone-enabled Raspberry Pi 3B+ and a RISC-V Keystone-based HiFive Unmatched Rev~B board, act as remote controlled endpoints.

\begin{figure}[!t]
\centering
\includegraphics[width=2.7in]{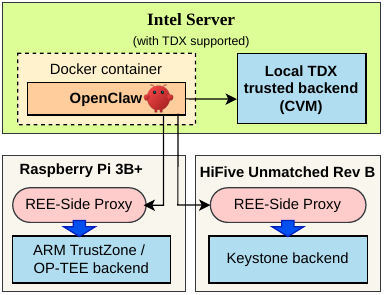}
\caption{Experimental deployment topology. OpenClaw runs on the Intel TDX server, where the local TDX backend performs primary trusted handling and produces audited control commands. The Raspberry Pi 3B+ and HiFive Unmatched Rev~B act as remote terminal targets that verify received TDX-audited commands through local TrustZone/OP-TEE or Keystone backends before constrained local execution.}
\label{fig:experimental_deployment}
\end{figure}

As shown in Fig.~\ref{fig:experimental_deployment}, OpenClaw runs inside a Docker container on the Intel server and cooperates with a local TDX trusted backend. For local operations, the protected path is handled by the TDX backend on the same server. For remote operations, OpenClaw sends TDX-audited commands to the terminal devices, where the local TrustZone/OP-TEE or Keystone backend verifies the command binding and integrity before the REE-side proxy dispatches it to the constrained local execution path.

This setup allows us to evaluate both local trusted-handling overhead and the additional overhead of delivering and verifying TDX-audited commands on heterogeneous remote terminals. The hardware and software configurations are summarized in Table~\ref{tab:experimental_platforms}. Unless otherwise stated, the baseline configuration corresponds to direct OpenClaw execution without the proposed protection mechanisms, while the protected configuration enables TDX-based trusted handling for local operations and remote trusted verification for TDX-audited commands.

\begin{table*}[!t]
\caption{Experimental platform configurations.}
\label{tab:experimental_platforms}
\centering
\renewcommand{\arraystretch}{1.12}
\begin{tabular}{p{2.0cm} p{2.2cm} p{3.5cm} p{2.2cm} p{2.4cm} p{2.8cm}}
\hline
\textbf{Role} & \textbf{Platform} & \textbf{CPU} & \textbf{Memory} & \textbf{Storage} & \textbf{Operating System} \\
\hline

\parbox[t]{2.0cm}{Host-side SHCUA\\platform}
&
\parbox[t]{2.2cm}{Intel TDX\\server}
&
\parbox[t]{3.5cm}{Intel Xeon Platinum 8558,\\48-Core @ 2.1\,GHz}
&
\parbox[t]{2.2cm}{256\,GB DDR5\\4800\,MT/s}
&
\parbox[t]{2.4cm}{Samsung SSD 990\\EVO Plus 2\,TB}
&
\parbox[t]{2.8cm}{Ubuntu 24.04}
\\

\parbox[t]{2.0cm}{Remote controlled\\target}
&
\parbox[t]{2.2cm}{Raspberry Pi\\3B+}
&
\parbox[t]{3.5cm}{Broadcom BCM2837B0,\\64-bit quad-core\\Cortex-A53 @ 1.4\,GHz}
&
\parbox[t]{2.2cm}{1\,GB LPDDR2\\SDRAM}
&
\parbox[t]{2.4cm}{64\,GB\\TF card}
&
\parbox[t]{2.8cm}{Raspbian Stretch\\(2018-11-13, for\\OP-TEE compatibility)}
\\

\parbox[t]{2.0cm}{Remote controlled\\target}
&
\parbox[t]{2.2cm}{HiFive Unmatched\\Rev~B}
&
\parbox[t]{3.5cm}{SiFive U74-MC,\\quad-core RISC-V\\@ 1.4\,GHz}
&
\parbox[t]{2.2cm}{16\,GB\\DDR4}
&
\parbox[t]{2.4cm}{31\,GB eMMC\\(SD32G)}
&
\parbox[t]{2.8cm}{Ubuntu 22.04\\LTS}
\\

\hline
\end{tabular}
\end{table*}

\subsubsection{Workload Configuration}

To evaluate both ordinary and security-critical SHCUA behavior, we use three representative OpenClaw-based workloads and define a fixed task list for baseline and protected configurations. The tasks cover local server operations, remote terminal operations, benign cases, security-critical cases, and both non-privileged and privileged execution paths.

We use the Django project repository as the ordinary local workspace, because it provides a well-known and representative project-style file structure, including documents, source files, tests, scripts, and project-local configuration-related files. This allows ordinary workloads to be instantiated over realistic engineering artifacts rather than ad hoc examples. For security-critical targets, we select representative Linux-sensitive files and configuration objects that are broadly present across systems, including \texttt{/etc/ssh/sshd\_config}, service definitions under \texttt{/etc/systemd/system/}, and user SSH materials under \texttt{~/.ssh/}.

The first workload is a \emph{document organization workload}, which represents ordinary non-privileged SHCUA usage over the Django-based workspace. The second workload is a \emph{protected configuration modification workload}, which covers both benign configuration updates and attempts to modify protected system configuration objects. The third workload is a \emph{command execution and export workload}, which covers benign bounded command execution as well as high-risk command patterns and sensitive cross-boundary export attempts.

The concrete task list used in the evaluation is summarized in Table~\ref{tab:workload_task_list}. This setup allows us to compare direct OpenClaw execution with the proposed protection architecture in terms of end-to-end task latency, operation-path overhead, denial behavior for unsafe requests, and practical usability.

\begin{table*}[!htbp]
\caption{Concrete workload task list used in the evaluation. \emph{Local} denotes operations on the Intel server running OpenClaw, while \emph{Remote} denotes operations issued to heterogeneous terminal devices through the protected remote-control path.}
\label{tab:workload_task_list}
\centering
\small
\renewcommand{\arraystretch}{1.12}
\begin{tabular}{p{0.9cm} p{2.8cm} p{5.6cm} p{1.4cm} p{1.8cm} p{3.4cm}}
\hline
\textbf{ID} & \textbf{Workload} & \textbf{Concrete task instance} & \textbf{Scope} & \textbf{Risk type} & \textbf{Main operations involved} \\
\hline

W1-1
& Document organization
& Read ordinary Django workspace files, including \texttt{README.rst}, selected documentation files under \texttt{docs/}, and selected test files under \texttt{tests/}, then generate a merged summary report under a local output directory.
& Local
& Benign
& Workspace file read, workspace file write \\

W1-2
& Document organization
& Reorganize part of the Django workspace by copying selected files from \texttt{docs/} and \texttt{tests/} into a new local review subdirectory and renaming them according to a fixed archive rule.
& Local
& Benign
& Workspace file copy, rename, directory write \\

W1-3
& Document organization
& Invoke a benign local helper command within the approved Django workspace scope, such as listing files under \texttt{docs/}, searching text in \texttt{README.rst}, or sorting extracted lines from a test file.
& Local
& Benign
& Local command execution, ordinary file access \\

W1-4
& Document organization
& Read ordinary system-information files from a remote Raspberry Pi or HiFive target through its REE-side proxy, such as \texttt{/etc/os-release}, \texttt{/proc/meminfo}, or \texttt{/proc/cpuinfo}, and produce a local summary report.
& Remote
& Benign
& Remote request, ordinary remote file read, local file write \\

W2-1
& Protected configuration modification
& Modify an ordinary local project-level configuration-related file in the Django workspace, such as \texttt{tox.ini} or \texttt{pyproject.toml}.
& Local
& Benign
& Workspace configuration write \\

W2-2
& Protected configuration modification
& Update an ordinary remote user-level configuration file through the proxy path, such as \texttt{~/.bashrc} or \texttt{~/.profile}.
& Remote
& Benign / conservatively blocked in remote prototype 
& Remote request, ordinary remote configuration write \\

W2-3
& Protected configuration modification
& Attempt to modify a protected local Linux configuration target, such as \texttt{/etc/ssh/sshd\_config} or a service definition under \texttt{/etc/systemd/system/}.
& Local
& Security-critical
& Protected file write, trusted authorization, possible denial \\

W2-4
& Protected configuration modification
& Attempt to modify a protected Linux configuration target on the Raspberry Pi or RISC-V device, such as \texttt{/etc/ssh/sshd\_config} or a protected service file under \texttt{/etc/systemd/system/}.
& Remote
& Security-critical
& Remote request, protected remote write, trusted authorization, possible denial \\

W3-1
& Command execution and export
& Execute an approved benign local Linux command within a bounded scope over the Django workspace, such as \texttt{grep -R "deprecated" docs/ tests/}, \texttt{find docs tests -type f}, or \texttt{head -n 20 README.rst}.
& Local
& Benign
& Local command execution \\

W3-2
& Command execution and export
& Execute a benign remote command on the Raspberry Pi or HiFive target through the proxy path, such as \texttt{uname -a}, \texttt{df -h}, or \texttt{cat /etc/os-release}.
& Remote
& Benign
& Remote request, benign remote command execution \\

W3-3
& Command execution and export
& Attempt to invoke a high-risk local command pattern, such as \texttt{curl <url> | sh}, \texttt{wget <url> -O - | bash}, or another unbounded command form.
& Local
& Security-critical
& High-risk command execution, trusted authorization, possible denial or notification \\

W3-4
& Command execution and export
& Attempt to package or export protected Linux-sensitive files across a trust boundary, such as user SSH materials under \texttt{~/.ssh/} or protected configuration files including \texttt{/etc/ssh/sshd\_config}.
& Remote
& Security-critical
& Protected file read, packaging, cross-boundary export, notification or denial \\

\hline
\end{tabular}
\end{table*}

\subsection{Performance Evaluation and Discussion}

\begin{figure}[!t]
\centering
\includegraphics[width=\linewidth]{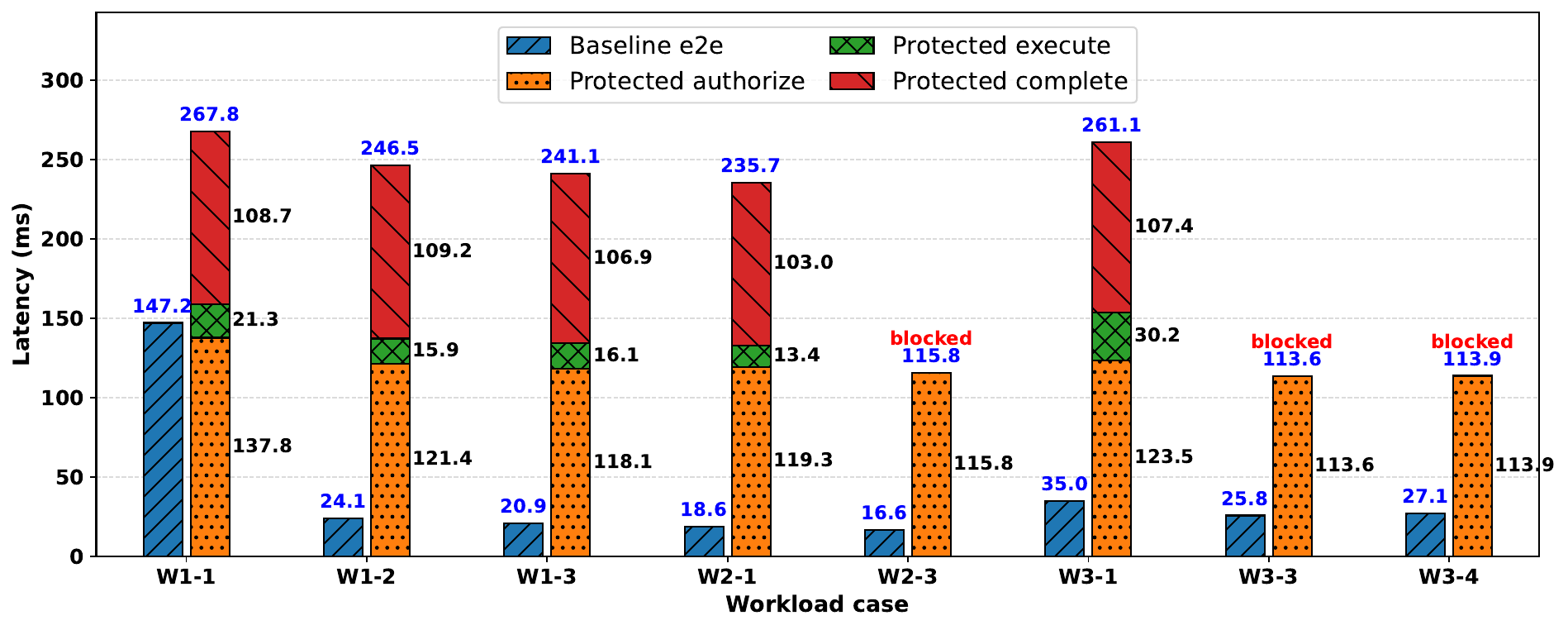}
\caption{Per-case latency breakdown for the local TDX evaluation. The baseline bar shows direct OpenClaw execution without trusted isolation, while the protected bar is decomposed into authorize, execute, and complete.}
\label{fig:perf_tdx}
\end{figure}

\begin{figure}[!t]
\centering
\includegraphics[width=\linewidth]{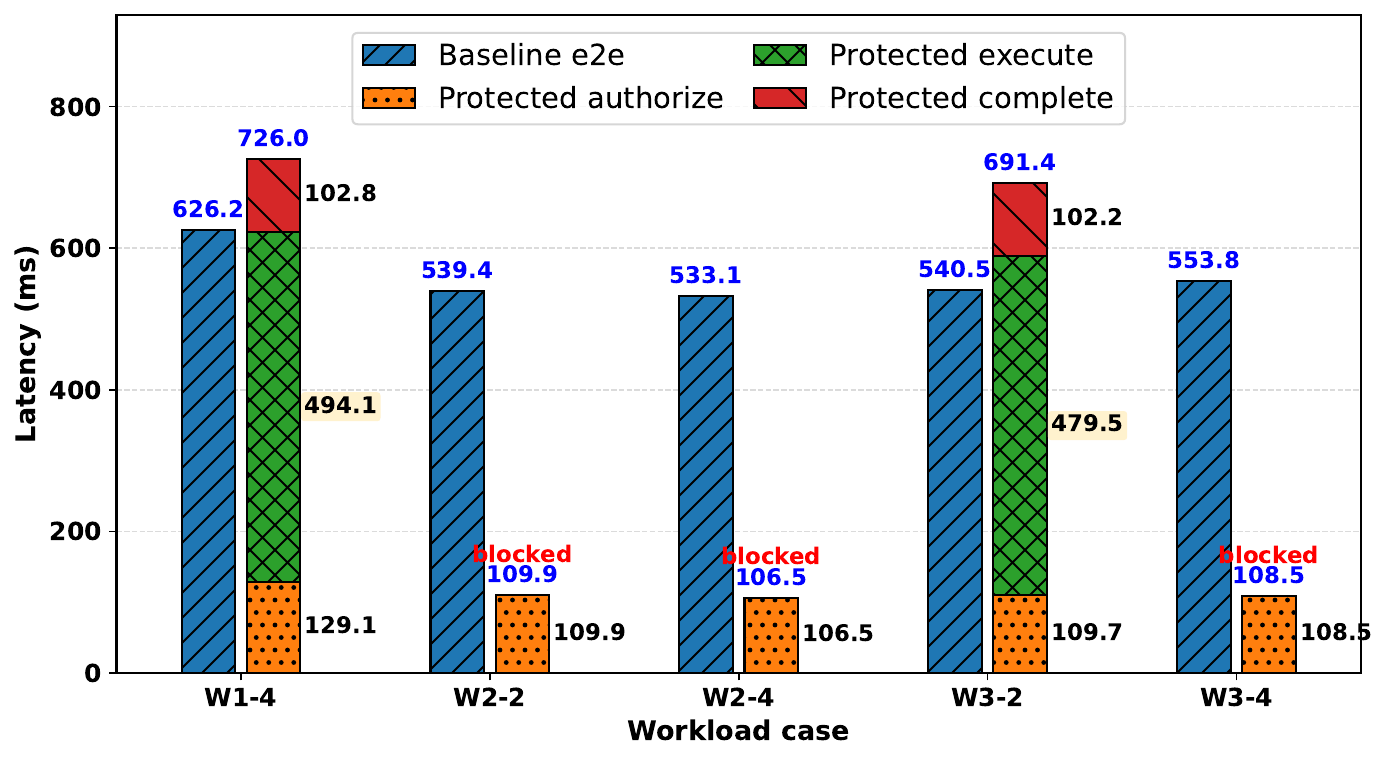}
\caption{Per-case latency breakdown for the remote OP-TEE evaluation. The measured cases correspond to OpenClaw-issued operations on the remote Raspberry Pi 3B+ target.}
\label{fig:perf_optee}
\end{figure}

\begin{figure}[!t]
\centering
\includegraphics[width=\linewidth]{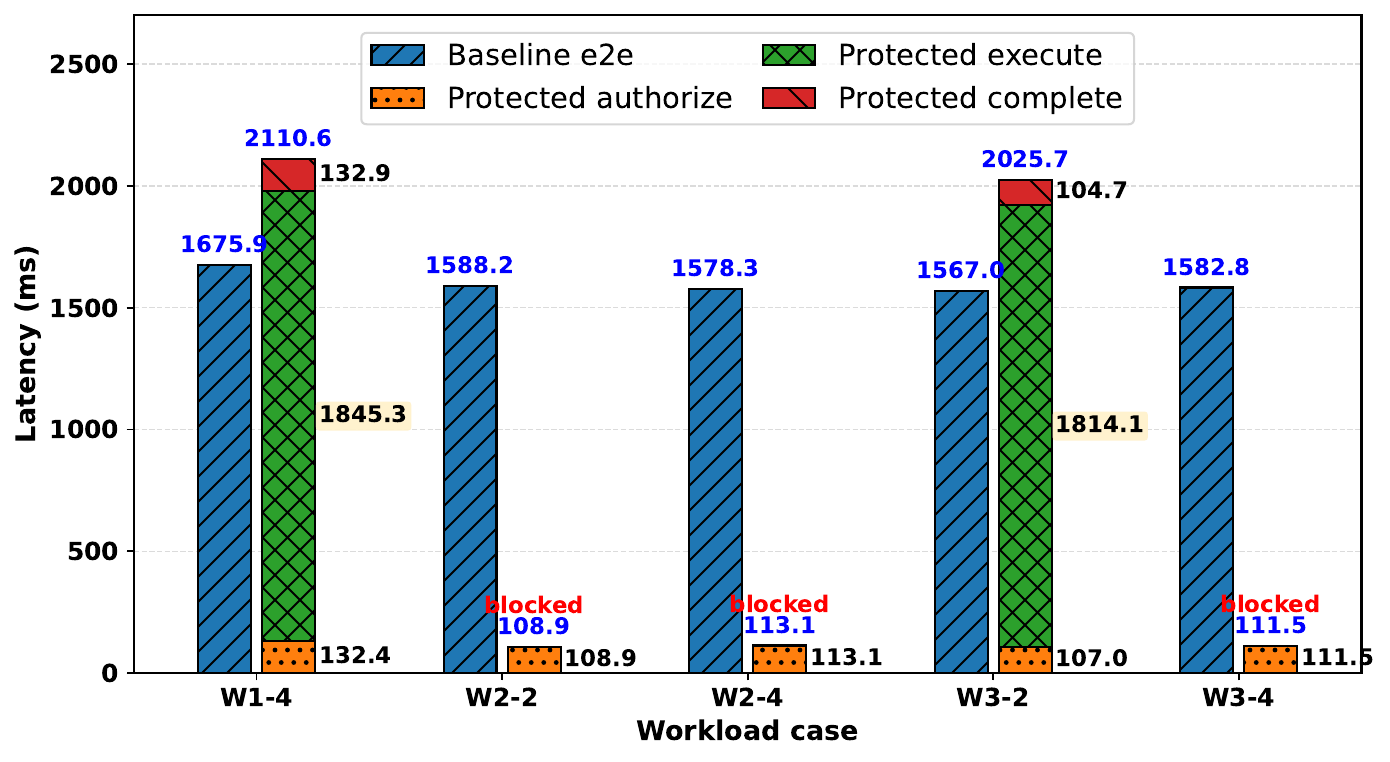}
\caption{Per-case latency breakdown for the remote Keystone evaluation. The measured cases correspond to OpenClaw-issued operations on the remote HiFive Unmatched target.}
\label{fig:perf_keystone}
\end{figure}

\subsubsection{Allowed Operations}

This subsection analyzes the workloads that are allowed to proceed under the protected configuration. These cases expose the full cost structure of the protected path, because the request is not stopped at authorization but continues into execution and completion. In our results, the allowed protected cases include W1-1, W1-2, W1-3, W2-1, and W3-1 in the local TDX setting, and W1-4 and W3-2 in the remote OP-TEE and Keystone settings.

Fig.~\ref{fig:perf_tdx}, Fig.~\ref{fig:perf_optee}, and Fig.~\ref{fig:perf_keystone} report the latency breakdown of these workloads under the baseline and protected configurations. Fig.~\ref{fig:perf_tdx} covers the local workload subset on the Intel TDX server, while Fig.~\ref{fig:perf_optee} and Fig.~\ref{fig:perf_keystone} cover the remote workload subset on the Raspberry Pi 3B+ and HiFive Unmatched targets.

For local TDX, the protected benign tasks W1-1, W1-2, W1-3, W2-1, and W3-1 complete in about 235.9--268.2~ms. Their \emph{execute} segments remain small, while most of the additional latency appears in \emph{authorize} and \emph{complete}. This is reasonable because the actual host-side actions in these workloads are lightweight local operations. Taking W1-1 as an example, the task mainly reads ordinary Django workspace files and writes a summary file. The baseline path therefore resembles ordinary local file processing. In the protected path, however, the same operation must be normalized into a trusted request, evaluated by the trusted operation plane, bound to an approved scope, and recorded through trusted evidence. The \emph{complete} phase further closes the result and evidence state. Thus, for local TDX workloads, the dominant cost comes from trusted control around a short local action rather than from the action itself.

For remote OP-TEE and Keystone, the allowed protected workloads are W1-4 and W3-2. These two cases are different from the local TDX workloads because the protected action must be delivered to a remote terminal and the result must be returned to the server. On OP-TEE, W1-4 and W3-2 have protected totals of 847.8~ms and 693.0~ms, with \emph{execute} contributing 494.1~ms and 479.5~ms. On Keystone, the same two cases reach 2238.4~ms and 2027.1~ms, with \emph{execute} contributing 1845.3~ms and 1814.1~ms. In these remote cases, the \emph{execute} phase includes command delivery, endpoint-side verification, target-side execution, and result return. This explains why the allowed remote workloads are dominated by \emph{execute}, whereas the local TDX workloads are dominated by authorization and completion around execution.

The latency difference between the OP-TEE and Keystone remote paths comes from both the remote-terminal hardware and the trusted-backend invocation mechanisms used in our prototype. For the OP-TEE path, the Raspberry Pi endpoint invokes a resident TrustZone secure-world service through the Linux TEE driver, the OP-TEE message protocol, and the ARM SMC calling convention~\cite{linux_optee_driver,optee_docs}. In our workload, this secure-side service is only used to verify the TDX-audited command and its authorization binding before the REE-side proxy performs the local action. Thus, the protected remote command follows a relatively short call path into an already initialized secure-world service.

The Keystone path is heavier in the current implementation. The HiFive endpoint uses a Keystone-style enclave backend, where trusted execution is mediated by a machine-mode security monitor responsible for PMP-based memory isolation, enclave entry and exit, attestation-related state, and system PMP synchronization~\cite{keystone_eurosys20,keystone_docs}. Keystone's execution model also involves enclave memory preparation and measurement; existing work on Keystone optimization reports that enclave startup may require hashing the executable memory image, and that statically linked enclave applications with runtime components increase the memory image size and startup cost~\cite{keystone_cache_2025}. Therefore, when an allowed remote command reaches the Keystone backend in our prototype, the measured \emph{execute} phase covers command transmission and target-side operation as well as enclave setup, memory measurement, runtime preparation, and monitor-mediated transitions. This explains why W1-4 and W3-2 are substantially slower on the HiFive/Keystone path than on the Raspberry Pi/OP-TEE path.

\subsubsection{Denied Operations}

This subsection analyzes rejected workloads, whose latency mainly reflects trusted classification, policy checking, and denial generation because they stop before host-side or remote target-side execution.

On local TDX, W2-3, W3-3, and W3-4 are rejected before execution, with protected totals of about 113.7--115.9~ms. Their \emph{execute} and \emph{complete} segments collapse to zero because no host-side action is dispatched after denial. The reason is that these workloads violate the protected policy at the operation-model level. For example, W2-3 has the action type of a write operation, but its target object is a protected local configuration file such as \texttt{/etc/ssh/sshd\_config}. In terms of the model in Section~III, the high object criticality and integrity impact raise the operation to a security-critical level, so the trusted operation plane returns denial rather than allowing the request to reach the constrained executor. W3-3 follows a similar pattern for high-risk command execution, while W3-4 is rejected because packaging or exporting protected files across a trust boundary creates a high confidentiality and externalization risk.

The remote denial cases follow the same logic but avoid an even larger cost. On OP-TEE, W2-2, W2-4, and W3-4 are denied at around 107.0--110.6~ms. On Keystone, the corresponding denial cases are around 109.5--113.6~ms. 
These values are lower than the allowed remote cases because the request is rejected before entering the command-delivery and target-side execution path.
Taking W2-4 as a representative example, the requested operation attempts to modify a protected configuration target on a remote terminal. Once the trusted authorization step determines that the operation exceeds the permitted remote scope, the system returns a denial without sending the command into the remote constrained execution path. This explains why W2-4 remains close to the authorization-side cost even though allowed remote workloads such as W1-4 and W3-2 incur much larger \emph{execute} latency.

W2-2 is a special case. Semantically, it is an ordinary remote user-configuration update, such as modifying \texttt{~/.bashrc} or \texttt{~/.profile}; however, the current prototype conservatively blocks remote configuration writes unless they are explicitly allowed by the remote policy profile. This choice reduces the risk of accidentally broadening remote write permissions during the prototype evaluation. Therefore, W2-2 should be interpreted as a conservative-denial case, not as a security-critical attack case or a measurement anomaly.

\subsubsection{Usability and Cost Interpretation}

The results show different user-visible costs for local server-side protection and remote terminal control. For local TDX workloads, allowed benign tasks remain within the sub-300~ms range. Because the protected host action is short, most added latency comes from the trusted control path, including classification, authorization, binding, and evidence closure. This overhead is still compatible with interactive SHCUA use, especially when \emph{complete} is treated as an asynchronous closure step after the protected action has taken effect.

For remote workloads, the cost pattern is different. Once a remote request is allowed, the total latency is dominated by command delivery, remote trusted verification, target-side execution, and result return. This explains why allowed remote cases are substantially slower than local TDX cases, while denied remote cases remain close to authorization-side latency because they do not enter the remote execution path. In practice, remote terminal control is therefore better suited to operations whose latency tolerance is higher or whose security benefit justifies the additional remote-control cost.

Overall, the evaluation confirms the intended enforcement behavior in the reported deployment: allowed benign operations can proceed, while unsafe or conservatively disallowed operations are blocked before execution. The dominant overhead is deployment-dependent. Local TDX workloads mainly reflect trusted-control overhead, whereas allowed remote workloads reflect the cost of the remote operation path; in our prototype, the HiFive/Keystone path further includes higher target-side overhead from the RISC-V enclave-based execution environment.

\subsection{Security Analysis}

The security of the proposed architecture relies on three properties: the SHCUA runtime is confined on the REE side, security-critical decisions are made inside the trusted operation plane, and each decision is bound to the host-side action that later consumes it. This section analyzes how these properties constrain host-level abuse under the threat model in Section~II.

\subsubsection{Trusted Decision Path}

The SHCUA runtime does not execute as an unrestricted host process. It runs in a constrained REE-side domain, and security-relevant accesses to protected files, commands, configuration objects, or communication sinks must pass through constrained executors. This prevents OpenClaw from directly exercising privileged host authority outside the protection path.

For security-critical operations, the decision point is moved out of ordinary REE logic. After an operation is extracted and normalized into a trusted request, the trusted operation plane reconstructs the corresponding operation instance, evaluates its risk factors, computes its security level, and derives the enforcement decision according to the model in Section~III. Therefore, a compromised REE component cannot by itself classify a sensitive operation as ordinary or issue a valid authorization for an arbitrary protected action.

The authorization returned by the trusted operation plane is scoped to the concrete request. It binds the action, target object, approved scope, context, sequence number, and freshness information. The constrained executor checks this binding before touching the protected resource. As a result, an approved operation cannot be legitimately broadened, redirected, or replayed as another protected action through the normal execution path.

\subsubsection{Pre-Execution Enforcement and Evidence Closure}

The protection path enforces security before protected host actions occur. In the evaluation, denied cases stop in the \emph{authorize} phase and never enter \emph{execute}. This shows that the mechanism is not a post-hoc audit layer: unsafe requests are blocked before they reach the corresponding host-side operation.

Allowed operations are executed only through constrained executors under the authorization returned by the trusted plane. The executor does not make the security decision; it applies the scoped decision to the pending host-side action. This separation keeps policy judgment inside the trusted domain while preserving compatibility with host-side resource access.

The \emph{complete} phase closes the request--decision--result chain. It records whether the authorized operation completed, failed, or was rejected, and binds the outcome to the trusted request and decision. When \emph{complete} is synchronous, the workflow returns only after evidence closure. When it is deferred, user-facing latency can be reduced, but evidence closure becomes asynchronous. In both cases, the security record remains tied to the trusted authorization path rather than to ordinary REE-side logs alone.

\subsubsection{Resistance to REE-Side Manipulation}

A compromised REE may drop requests, delay forwarding, suppress completion, or refuse to execute an authorized action. These behaviors can cause denial of service or workflow disruption, but they do not create a valid authorization for a different protected operation.

The key reason is that the REE does not control the trusted decision path. It cannot forge a valid authorization without the trusted operation plane, and it cannot reuse an authorization unless the action, object, scope, context, sequence number, and freshness checks match. Argument substitution, target redirection, stale replay, or cross-operation reuse therefore either fails at the constrained executor or creates an inconsistency in the request--decision--result evidence chain.

This is the intended security boundary. The architecture does not make the REE trustworthy, and it does not prevent all forms of disruption. It prevents REE-side manipulation from silently becoming legitimate protected execution under a different meaning.

\subsubsection{Remote Terminal Verification}

In our prototype, Intel TDX is the primary backend that hosts the trusted operation plane. It performs server-side classification, authorization, binding, and evidence generation. The OP-TEE and Keystone components used in the remote experiments have a narrower role: they verify that a received remote command is bound to the TDX-audited authorization and has not been tampered with before constrained local execution.

Thus, the remote setting does not assume that every terminal-side TEE hosts an equivalent trusted operation plane. Security-critical decisions are anchored in the TDX-backed trusted path, while terminal-side TEEs act as lightweight verification anchors for received commands. This matches the deployment model used in the evaluation and avoids relying on resource-constrained embedded TEEs to support full OpenClaw-side trusted control.

\subsubsection{Summary}

The architecture constrains SHCUA host-level abuse by combining constrained REE-side deployment, TEE-backed trusted decisions, scoped execution through constrained executors, and evidence closure. A compromised REE can still disrupt the workflow, but it cannot validly authorize, broaden, redirect, replay, or silently complete a protected operation outside the trusted decision path. In remote terminal scenarios, the same security semantics are preserved by verifying TDX-audited commands before local constrained execution, rather than by duplicating the full trusted operation plane on each terminal.

\section{Related Work}

\subsection{SHCUA Security}

Self-hosted computer-use agents (SHCUAs) extend conventional language-model assistants from text generation to direct interaction with host-side resources, including files, commands, browsers, plugins, and communication channels. This shift changes the security problem from unsafe responses to unsafe host-level actions. Public OpenClaw-related incidents have already exposed several concrete attack surfaces, including website-to-local-agent takeover, authentication-token-related remote code execution, malicious skill abuse, and malware delivery through fake installers~\cite{openclaw2026repo,oasis2026clawjacked,github2026openclawrce,immersive2026skills,malwarebytes2026fakeinstaller}. These cases show that the agent's legitimate capability surface can itself become the attack target once it is connected to real tools and resources.

Recent research has started to analyze this class of systems more systematically. Existing studies examine OpenClaw and computer-use agents from the perspectives of system-level attack surfaces, harmful action benchmarks, visual prompt injection, action traceability, and agent-specific threat taxonomies~\cite{foerster2026camels,wang2026agentasset,wang2026systematicopenclaw,cao2025vpibench,peng2026agenttrace,shan2026clawgrip,ying2026threatsdefensesopenclaw,lu2026clawless,zhan2024injecagent,debenedetti2024agentdojo,zhang2024asb,kuntz2025osharm,evtimov2025wasp,feng2026agenthazard,wei2026clawsafety,luo2025codeagenthacker}. These works are valuable because they make the threat landscape concrete: indirect prompt injection, adversarial visual content, unsafe tool invocation, over-privileged execution, and weak action traceability all become realistic concerns in SHCUA deployments. Their main focus, however, is attack discovery, taxonomy construction, or benchmark evaluation. They do not provide a trusted runtime architecture that determines whether a concrete host operation should be allowed, denied, elevated, or audited under a stronger trust boundary.

\subsection{Protection Mechanisms for Agents}

Existing protection mechanisms for agents can be grouped into three broad directions. The first direction operates before or during model inference. Representative work studies prompt-injection filtering, input separation, prompt-leakage mitigation, model-side safety behavior, memory poisoning, and trust-aware retrieval~\cite{wang2026icon,alizadeh2025simplepromptleakage,wang2026landscapepromptinjectionthreats,sunil2026memorypoisoning}. These mechanisms can reduce the probability that malicious content influences the agent's decision process. They are less suited to deciding whether a concrete host-side operation is safe after the agent has already produced a tool call.

The second direction introduces policy and runtime enforcement for agents. Examples include runtime policy specifications, policy compilation, layered guardrails, runtime governance, pre-action authorization, probabilistic runtime checking, behavioral contracts, and privilege-governed execution~\cite{wang2025agentspec,palumbo2026pcas,wang2025mi9,koch2026layeredguardrails,kaptein2026runtimegovernance,uchibeke2026preactionauth,shi2025progent,wang2025pro2guard,bhardwaj2026agentbehavioralcontracts,zhang2026grantbox}. This line is closer to our work because it treats agent execution as a controlled runtime process. It also highlights the importance of explicit policy, action mediation, and permission-aware execution. The remaining gap is the trustworthiness of the enforcement path itself. If classification, authorization, scope binding, and audit generation are all performed by ordinary REE-side software, then a compromised or manipulated runtime may still misclassify operations, alter parameters after authorization, or produce unreliable evidence.

The third direction uses sandboxing, constrained execution, or bounded tool environments to reduce the agent's direct authority over host resources~\cite{wang2026safeclawr,foerster2026camels,li2026securityconsiderationsagents}. This principle is important for SHCUA deployments and is also used in our design. Constrained execution limits what the agent can directly reach, but it does not by itself determine who can be trusted to classify a security-critical operation, bind the approved scope, trigger user confirmation, or generate verifiable evidence. Our work combines constrained REE-side execution with a TEE-backed trusted operation plane so that security-critical decisions are not left solely to the host runtime.

\subsection{System-Level Protection Foundations}

The system-security literature provides several foundations for our design. Privilege separation and enforceable security policies argue that security-sensitive operations should be mediated by smaller, more controllable components rather than by ordinary application logic~\cite{provos2003privsep,schneider2000enforceable}. This idea is reflected in our split between the ordinary OpenClaw runtime and REE-side constrained executors.

Trusted paths and evidence-oriented auditing provide another relevant foundation. Trusted-path guidance emphasizes that security-critical interactions should be bound to a trustworthy communication and control path, while evidence-based audit stresses that audit value depends on whether evidence semantics are generated and preserved reliably~\cite{nist2020trustedpath,vaughan2008evidenceaudit}. These ideas are especially important for SHCUAs because ordinary host logs may be suppressed, reordered, or forged if the REE-side runtime is compromised. Our design applies this principle to agent operations by placing classification, authorization, binding, and evidence generation in a TEE-backed trusted operation plane.

Recent system-level studies of agentic computing also argue that agent security should be treated as a runtime and systems problem rather than only a prompt-engineering problem~\cite{christodorescu2025systemssecurityagentic,deng2024agentsunderthreat,wang2025agentxploit,chen2024agentpoison,zou2026poisononce,maloyan2026promptinjectioncodingassistants}. These works motivate action mediation, provenance, control boundaries, and runtime structure. Our work follows this direction in the specific setting of SHCUAs, where the agent has direct host-side operational capability and where security decisions must be tied to concrete operations over real resources.

\subsection{Comparison with Existing Work}

Table~\ref{tab:related_work_comparison} summarizes the relationship between representative prior work and our approach. Existing SHCUA and OpenClaw studies identify realistic attacks and evaluation tasks, but usually stop at analysis or benchmarking. Input- and model-side defenses reduce adversarial influence before execution, but they do not provide a trusted basis for host-side operation decisions. Policy and runtime enforcement frameworks constrain agent actions, but their enforcement logic is usually still hosted in the ordinary runtime. General system-level protection schemes provide useful principles such as privilege separation, trusted paths, and evidence-based auditing, but they are not designed around operation-centric SHCUA control.

Our work combines these ideas in a cloud-native SHCUA protection architecture. It models concrete operations as security-relevant instances, evaluates their criticality through operation-level factors, protects classification and authorization in a TEE-backed trusted operation plane, and uses constrained executors for bounded host-resource access. In the remote-terminal setting, the server-side TDX path produces audited commands, while terminal-side trusted components verify the received command before constrained local execution. This makes the contribution different from both attack-taxonomy papers and ordinary host-side policy engines.

\begin{table*}[!t]
\caption{Comparison with representative existing work.}
\label{tab:related_work_comparison}
\centering
\small
\renewcommand{\arraystretch}{1.12}
\begin{tabular}{p{3.1cm} p{3.6cm} p{2.1cm} p{2.1cm} p{2.6cm} p{2.5cm}}
\hline
\textbf{Work category} &
\textbf{Representative works} &
\textbf{Targets SHCUA host-level abuse} &
\textbf{Operation-level risk modeling} &
\textbf{Trusted classification / decision path} &
\textbf{Remote terminal verification} \\
\hline

SHCUA/OpenClaw security analysis
& \cite{wang2026agentasset,wang2026systematicopenclaw,shan2026clawgrip,wei2026clawsafety}
& Yes
& Partial
& No
& No \\

Computer-use agent attack benchmarks
& \cite{cao2025vpibench,zhan2024injecagent,debenedetti2024agentdojo,kuntz2025osharm,feng2026agenthazard}
& Mostly yes
& Partial
& No
& No \\

Input/model-side defenses
& \cite{wang2026icon,alizadeh2025simplepromptleakage,wang2026landscapepromptinjectionthreats,sunil2026memorypoisoning}
& Partial
& No
& No
& No \\

Policy/runtime enforcement for agents
& \cite{wang2025agentspec,palumbo2026pcas,wang2025mi9,kaptein2026runtimegovernance,uchibeke2026preactionauth,shi2025progent,wang2025pro2guard,bhardwaj2026agentbehavioralcontracts}
& Partial
& Partial
& Usually no
& No \\

Sandboxing and constrained execution
& \cite{wang2026safeclawr,foerster2026camels,li2026securityconsiderationsagents}
& Partial
& No / partial
& No
& No \\

General system-level protection
& \cite{provos2003privsep,vaughan2008evidenceaudit,schneider2000enforceable,nist2020trustedpath}
& No
& No
& Partial
& No \\

\textbf{This work}
& --
& \textbf{Yes}
& \textbf{Yes}
& \textbf{Yes}
& \textbf{Yes} \\

\hline
\end{tabular}
\end{table*}

\section{Conclusion}

Self-hosted computer-use agents create a new host-level abuse surface by translating natural-language intent into concrete operations over real system resources. This paper addressed this problem through an operation-centric security model and a risk-driven minimal-confinement strategy. The proposed design keeps ordinary SHCUA functionality on the constrained REE path, while anchoring security-critical classification, authorization, binding, audit evidence, notification, and selected execution-control decisions in a cloud-native TEE-backed trusted operation plane.

We instantiated this architecture using OpenClaw as a representative SHCUA and Intel TDX as the primary trusted backend. The evaluation shows that the design can block unsafe or policy-disallowed operations before execution, preserve ordinary functionality for allowed workloads, and provide trusted evidence for security-relevant actions. The remote experiments further show how TDX-audited commands can be verified by terminal-side trusted components before constrained local execution. Overall, the proposed approach provides a practical path toward auditable, verifiable, and selectively constrained SHCUA operation without moving the entire agent stack into trusted isolation.

\section*{Acknowledgments}
The authors would like to thank the editor-in-chief, associate editor, and reviewers for their valuable comments and suggestions. This research was supported by the National Natural Science Foundation of China (62232013, U24A20243, 62572377, 62302363), the Innovation Capability Support Program of Shaanxi (No. 2023-CX-TD-02), the Xidian University Specially Funded Project for Interdisciplinary Exploration (No. TZJHF202502) and the Fundamental Research Funds for the Central Universities (No. ZDRC2202).


\bibliographystyle{IEEEtran}
\bibliography{refs}

\begin{IEEEbiography}[{\includegraphics[width=1in,height=1.25in,clip,keepaspectratio]{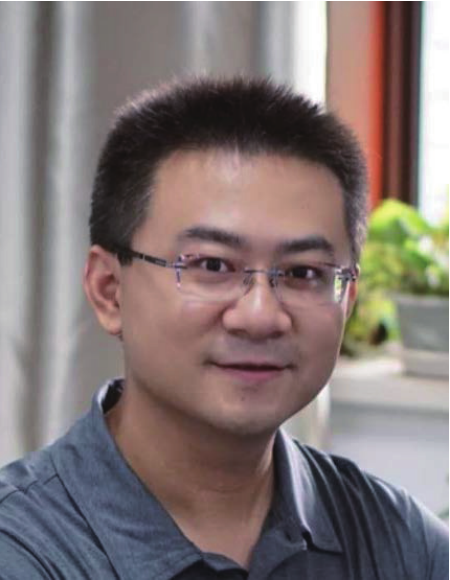}}]{Di Lu}
(Member, IEEE) received the B.S., M.S., and Ph.D. degrees in computer science and technology from Xidian University, China, in 2006, 2009, and 2014. Now he is a full  Professor in the School of Computer Science and Technology at Xidian University. His research interests include trusted computing, confidential computing, system and network security.
\end{IEEEbiography}

\begin{IEEEbiography}[{\includegraphics[width=1in,height=1.25in,clip,keepaspectratio]{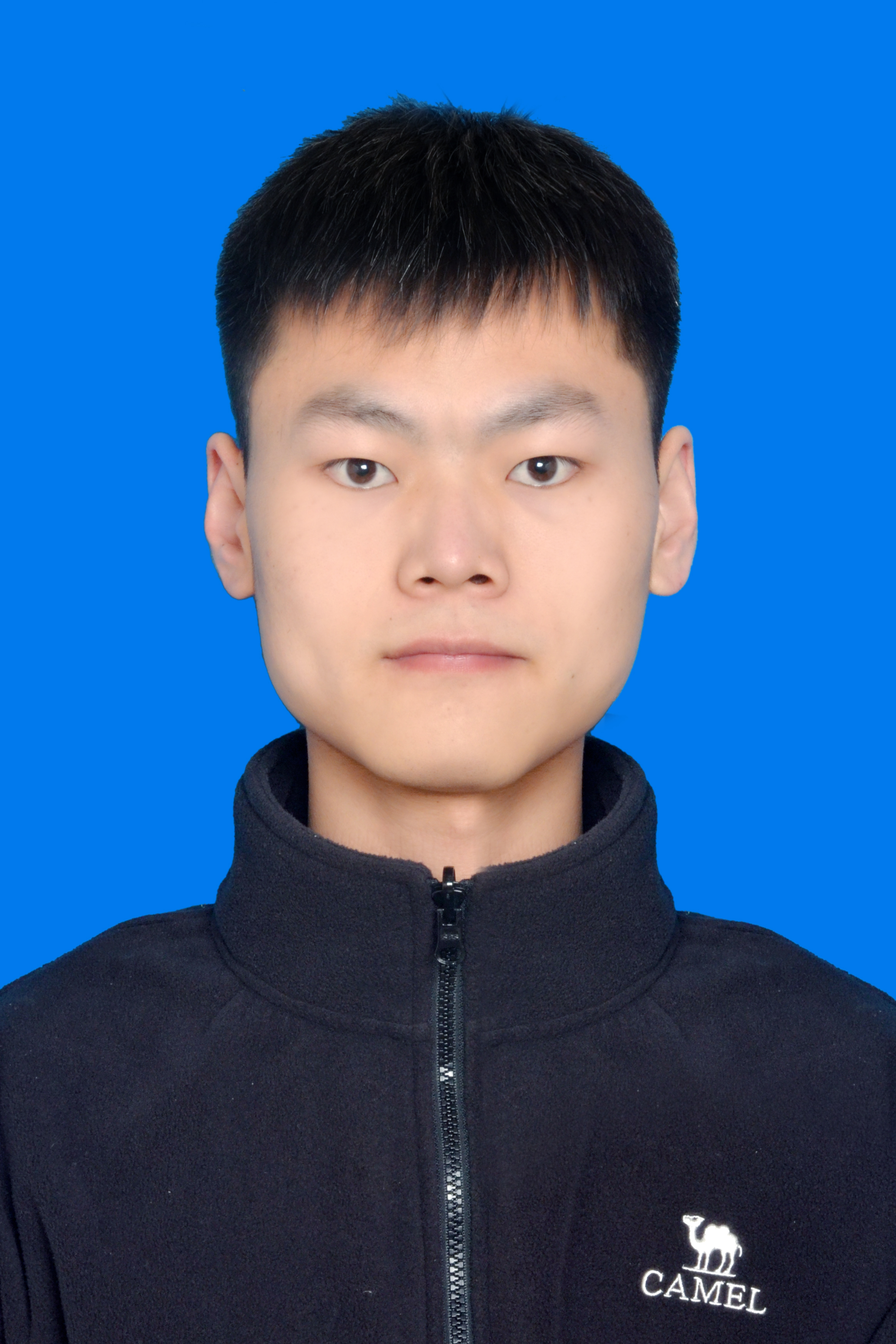}}]{Bo Zhang}
received the BS degree from Nanyang Institute of Technology, China, in 2025. He is currently pursuing an MS degree in the School of Computer Science and Technology at Xidian University, China. His research interests include trusted computing and embedded system security.
\end{IEEEbiography}

\begin{IEEEbiography}[{\includegraphics[width=1in,height=1.25in,clip,keepaspectratio]{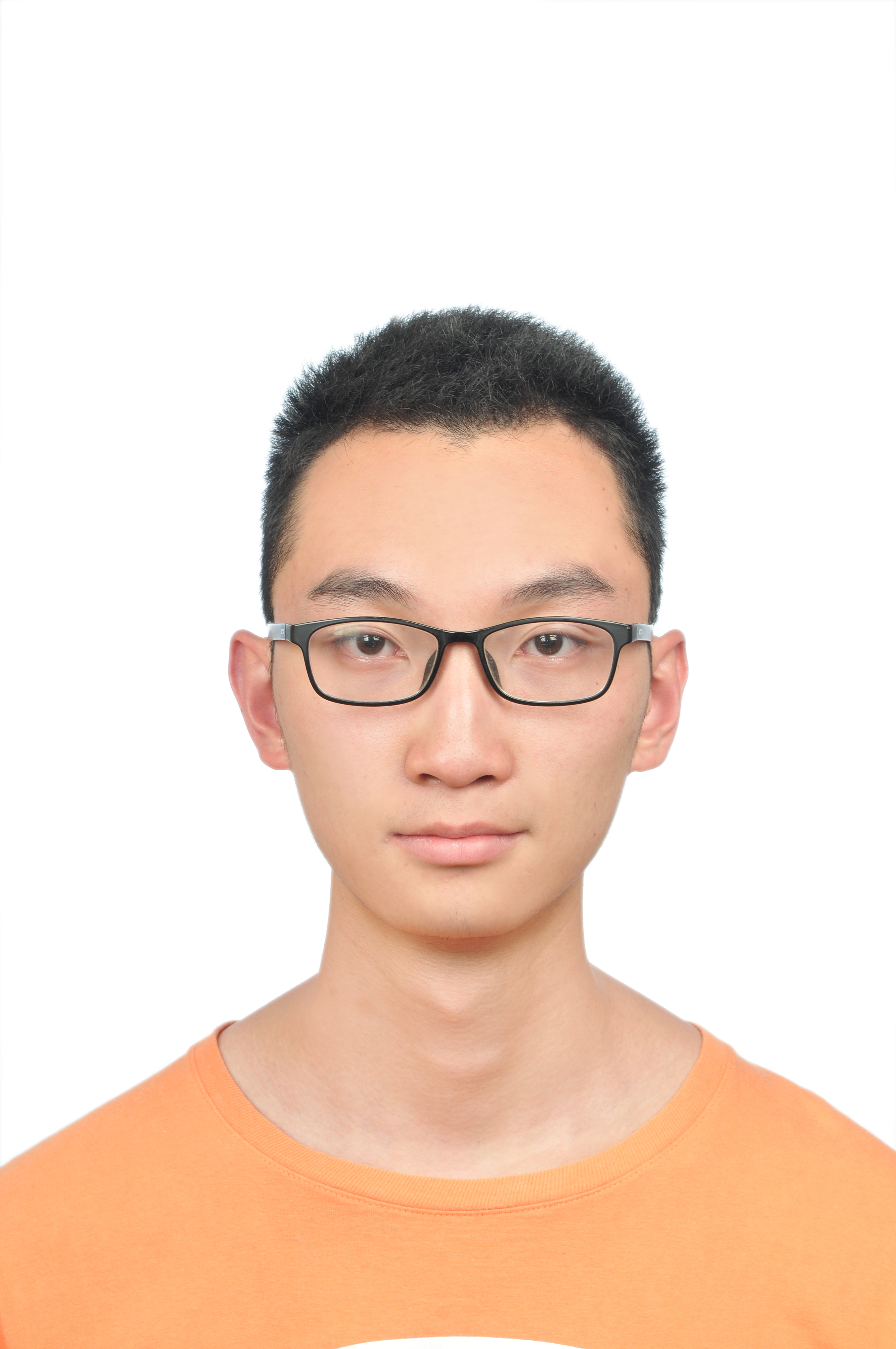}}]{Xiyuan Li} is currently pursuing the BS degree in computer science and technology with the School of Computer Science and Technology, Xidian University, China. His research interests include TEE technology and secure agent systems. \end{IEEEbiography}

\begin{IEEEbiography}[{\includegraphics[width=1in,height=1.25in,clip,keepaspectratio]{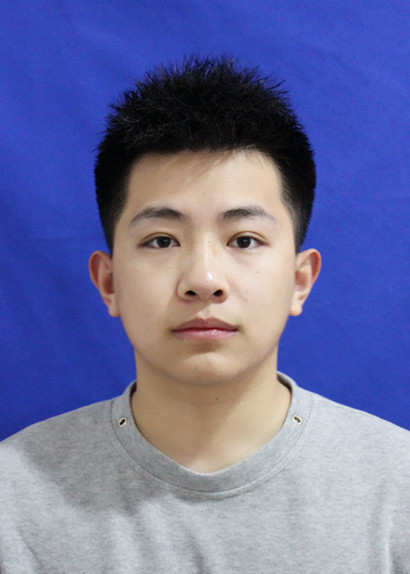}}]{Yongzhi Liao}
received the BS degree from Xi'an Shiyou University, China, in 2024. He is currently pursuing an MS degree in the School of Computer Science and Technology at Xidian University, China. His research interests include LLM security and 3D reconstruction security.
\end{IEEEbiography}

\begin{IEEEbiography}[{\includegraphics[width=1in,height=1.25in,clip,keepaspectratio]{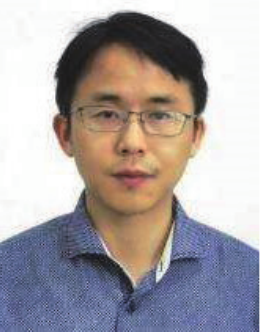}}]{Xuewen Dong}
(Member, IEEE) received the B.E., M.S., and Ph.D. degrees in computer science and technology from Xidian University, Xi’an, China, in 2003, 2006, and 2011, respectively. From 2016 to 2017, he was a Visiting Scholar with Oklahoma State University, Stillwater, OK, USA. Currently, he is a Professor with the School of Computer Science and Technology, Xidian University. His research interests include blockchain and the security of smart systems.
\end{IEEEbiography}

\begin{IEEEbiography}[{\includegraphics[width=1in,height=1.25in,clip,keepaspectratio]{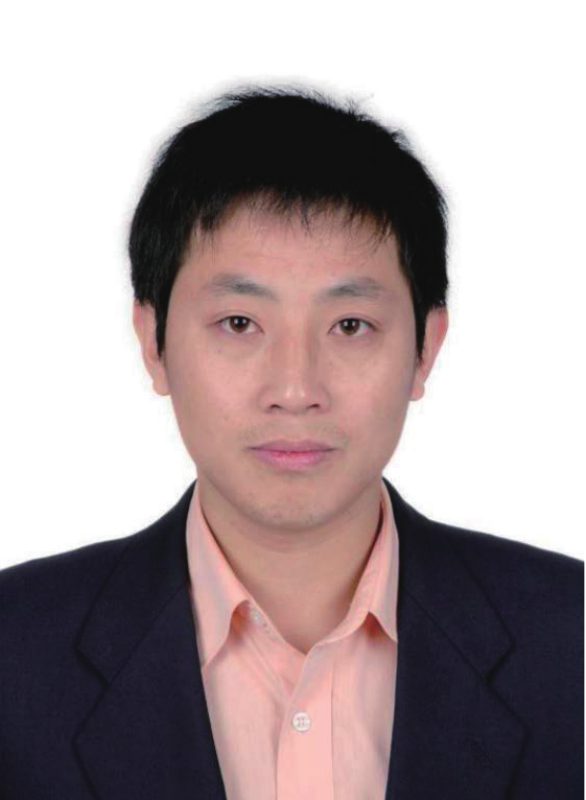}}]{YuLong Shen}
(Member, IEEE) received the BS and MS degrees in computer science and a PhD degree in cryptography from Xidian University, Xi’an, China, in 2002, 2005, and 2008, respectively. He is currently a professor with the School of Computer Science and Technology, Xidian University. His research interests
include wireless network security and cloud computing security.
\end{IEEEbiography}

\begin{IEEEbiography}[{\includegraphics[width=1in,height=1.25in,clip,keepaspectratio]{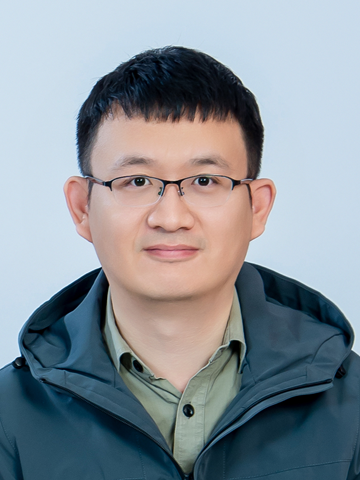}}]{Zhiquan Liu}
received the B.S. degree from the School of Science, Xidian University, Xi'an, China, in 2012, and the Ph.D. degree from the School of Computer Science and Technology, Xidian University, Xi'an, China, in 2017.

He is currently a full professor, doctoral supervisor, and deputy dean with the College of Cyber Security, Jinan University, Guangzhou, China. His current research focuses on security, trust, privacy, and intelligence in vehicular networks and UAV networks. He currently serves as the area editor or associate editor of multiple SCI-index journals, such as IEEE TIFS, IEEE TDSC, IEEE TII, IEEE TVT, IEEE IOTJ, IEEE Network, Information Fusion, etc. His homepage is https://www.zqliu.com.
\end{IEEEbiography}

\begin{IEEEbiography}[{\includegraphics[width=1in,height=1.25in,clip,keepaspectratio]{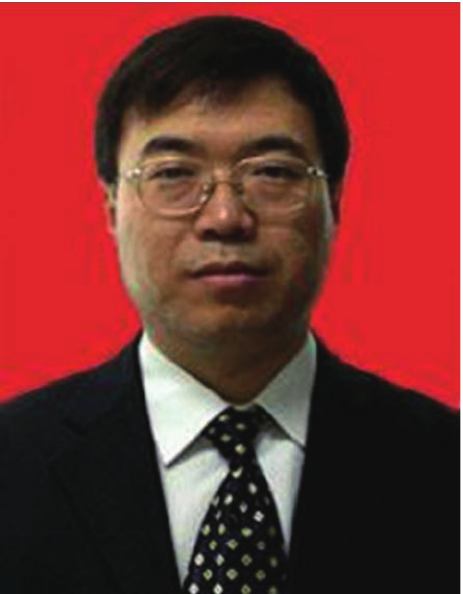}}]{Jianfeng Ma}
(Member, IEEE) received the BS degree in mathematics from Shaanxi Normal University, China, in 1985, and the MS and PhD degrees in computer software and communications engineering from Xidian University, China, in 1988 and 1995, respectively. Now, he is a professor with the School of Cyber Engineering, Xidian University, China. His current research interests include distributed systems, computer networks, and information and network security.
\end{IEEEbiography}

\end{document}